# Dramatic changes in DNA Conductance with stretching: Structural Polymorphism at a critical extension

Saientan Bag[1], Santosh Mogurampelly[1#], William A Goddard[2] and Prabal K Maiti[1*]

[1]Center for Condensed Matter Theory, Department of Physics, Indian Institute of Science, Bangalore 560012, India

[2]IISc-DST Centenary Chair Professor, Department of Physics, Indian Institute of Science, Bangalore 560012, India and Materials and Process Simulation Center, California Institute of Technology, Pasadena, California 91125, USA

*Abstract*— **In order to interpret recent experimental studies of the dependence of conductance of ds-DNA as the DNA is pulled from the 3'end1-3'end2 ends, which find a sharp conductance jump for a very short (4.5 %) stretching length, we carried out multiscale modeling, to predict the conductance of dsDNA as it is mechanically stretched to promote various structural polymorphisms. We calculate the current along the stretched DNA using a combination of molecular dynamics simulations, non-equilibrium pulling simulations, quantum mechanics calculations, and kinetic Monte Carlo simulations. For 5'end1-5'end2 attachments we find an abrupt jump in the current within a very short stretching length (6 Å or 17 %) leading to a melted DNA state. In contrast, for 3'end1-3'end2 pulling it takes almost 32 Å (84 %) of stretching to cause a similar jump in the current. Thus, we demonstrate that charge transport in DNA can occur over stretching lengths of several nanometers. We find that this unexpected behaviour in the B to S conformational DNA transition arises from highly inclined base pair geometries that result from this pulling protocol. We find that the dramatically different conductance behaviors for two different pulling protocols arise from the nature of how hydrogen bonds of DNA base pairs break.**

*[*maiti@physics.iisc.ernet.in](mailto:maiti@physics.iisc.ernet.in)

[#] Present Address: Department of Chemical Engineering, The University of Texas at Austin, Austin, Texas 78712, United States.

# INTRODUCTION

Understanding how charge transport in dsDNA depends upon its environment is important for designing such applications as bio sensors or nanowires [1-5] and it is important to understanding oxidative damage [6-11] and DNA repair. The charge transport properties of DNA have previously been reported to be that of an insulator[12, 13], a conductor[14-16], a semiconductor, and even a superconductor[17]. This Lack of reproducibility and the strong dependence on the external environment have made the subject of DNA conductance quite controversial. The Charge transport in DNA is mediated by $\pi - \pi$ stacking interaction (coupling) through its bases, which in turn is controlled by the rise and twist of the bases and by the external environment[18-20]. Consequently to obtain reproducible DNA conductance measurements, the experiments necessarily require a reproducible electronic coupling between the bases, which is difficult to achieve experimentally [21-23]. Sophisticated single molecule experiments on DNA by Xu et al.[19], Legrand et al.[24] and Song et al.[25] have concluded that [26] DNA is a semiconductor under ambient conditions. In ionic solutions in which DNA is able to keep its native state, DNA mainly exhibits semiconducting properties. In these experiments [19, 24, 25] the DNA was kept in an ionic environment between two electrodes and the current was measured either by scanning tunneling microscopy or by the break junction technique.

Extensive theoretical calculations have been reported on the charge transport of DNA. Cramer et al.[27] used the Marcus–Hush formalism to calculate the I-V characteristics of the DNA where the Marcus-Hush parameter was calculated using the extended Su-Schrieffer-Heeger Hamiltonian. Sairam et al.[18] combined MD simulations using a force field with first principles quantum mechanics (QM) calculations to calculate transmission coefficient for charge transport through DNA.

Recently Tao et al.[28] studied the dependence of the conductance of ds-DNA as DNA is pulled from the 3'end1-3'end2 ends. They observed a sharp conductance jump for a very short (4.5 %) stretching length, which they attributed to the breaking of hydrogen bonds between the terminal base-pairs at the DNA termini. In a related work, they compared the critical stretching length (length at which the conduction jump occurs) of various closed-end and free end DNA systems[29]. Tao et al.[30] also studied the molecular conductance and piezoresistivity of dsDNA for various lengths and sequences. In a very recent work, Joshua et al.[31] reported the increase in conductance of the DNA during its conformational change from B to A form.

In addition to the stretching of DNA from 3'end1-3'end2 as studied by Tao et al., there are primarily three other modes to stretch DNA, leading to very different kinds of structures [32, 33] depending on the stretching protocol. This raises the question of how the conductance in dsDNA depends on the mode of mechanical stretching.

In this paper we report a semiclassical Marcus-Hush type calculation of the charge transport through the dsDNA bases and present the current vs stretching length behavior for each of the four pulling protocols considered in our simulations. Our calculations combine MD simulations, non-equilibrium pulling simulations, QM calculations, and kinetic Monte Carlo simulations. In all the four cases, the jump in current was seen as the DNA is stretched but at different stretching lengths. The stretching length at which the jump in current occurs is defined as the critical stretching length $l_c$. For the case of 5'end1-5'end2 pulling, we find a short $l_c$ of 6 $\mathring{A}$, leading to a melted DNA state, while for 3'end1-3'end2 pulling, we find a high $l_c$ of 32 $\mathring{A}$, before the transformation from B to S-DNA. The $l_c$ for other two protocols (3'end1-5'end2 and 3'end1-5'end1) was found to have intermediate values.

The paper is organized as follows: in next section we describe the all atom MD simulations and the non-equilibrium pulling simulations which were used to predict the structures arising from mechanical pulling. In the methodology section we describe the Marcus-Hush formalism and the calculation scheme for the current through the DNA kept between two electrodes. The results are described in section III. In section IV we summarize the results.

## Simulation Details

The initial structures of the duplex DNA was generated using the *nucgen* module available in the AMBER suite of [34] programs. We then simulated dsDNA with lengths of 12, 14, 16, 18 and 20 base pairs, using the following sequences:

- d(CGCGAATTCGCG),
- d(CGCGAAATTTCGCG),
- d(CGCGAAAATTTTCGCG),
- d(CGCGAAAAATTTTTCGCG) and
- d(CGCGAAAAAATTTTTTCGCG).

Each dsDNA structure was solvated in a box of water using TIP3P water model. The water box dimensions were chosen to ensure 10 $\mathring{A}$ solvation of the dsDNA in each direction, when the DNA is fully stretched. Sufficient Na$^+$ counterions were added to neutralize the negative charge of phosphate backbone groups of the DNA. We followed the simulation protocol described in references [35, 36] to equilibrate the system at a temperature of 300 K and a pressure of 1 bar. For constant force pulling simulations, we used modified-SANDER,[34] previously developed in our group [37] to add the necessary forces at the dsDNA terminus. The external force was applied on the dsDNA using four different protocols as shown in Figure 1. Staring from 0 pN, the external force increased linearly with time at the rate of 0.0001 pN fs$^{-1}$. The trajectory file was saved every 2 ps.

**Methodology**

The experimental situation that we want to mimic with our conductance calculations is shown schematically in Figure 1(b). To measure the current through the dsDNA, we follow the semiclassical Marcus-Hush formalism[38, 39] in which charge transport is described as incoherent hopping of charge carriers between dsDNA bases. In this formalism, the charge transfer rate $\omega_{ij}$ from $i^{\text{th}}$ charge hopping site to the $k^{\text{th}}$ hopping site is given by

$$\omega_{ik} = \frac{|J_{ik}|^2}{\hbar} \sqrt{\frac{\pi}{\lambda k_B T}} \exp\left[-\frac{(\Delta G_{ik} - \lambda)^2}{4\lambda k_B T}\right]. \qquad (1)$$

where $J_{ik}$ is the transfer integral, defined as

$$J_{ik} = \langle \phi^i | \widehat{H} | \phi^k \rangle. \qquad (2)$$

Here $\phi^i$ and $\phi^k$ are the diabatic wave functions localized on the $i^{\text{th}}$ and $k^{\text{th}}$ sites, respectively. $\widehat{H}$ is the Hamiltonian for the two site system between which the charge transfer takes place, $\Delta G_{ik}$ is the free energy difference between the two sites, $\lambda$ is the reorganization energy, $\hbar$ is the Plank's constant, $k_B$ is the Boltzmann constant, and T is the absolute temperature.

In order to predict the dsDNA structures resulting from mechanical pulling, we performed non-equilibrium MD simulations as described in Section-I. Once the dsDNA structures are known, we remove the dsDNA backbone for additional calculations since we assume that the charge transport in DNA occurs solely through interaction of the stacked bases. Previous theoretical and experimental investigation have demonstrated that the charge transport in DNA is mediated by stacked nucleobases through strong π-π interaction [4, 20, 40, 41]. So we do not include backbone during the optimization. Then we calculate [42-44] the charge transfer rates between the neighboring bases between which the charge transport takes place. For instance if the electron is on the $i^{th}$ base, the charge can hop to any one of its five neighboring bases requiring the calculation of five different hopping terms. With these rates in hand, we perform kinetic Monte Carlo simulations to obtain the numerical value for the current. We used Density Functional Theory (DFT) to calculate the various terms appearing in the rate expression. The highest occupied molecular orbital (HOMO) and lowest unoccupied molecular orbital (LUMO) are used [45, 46] as diabatic wave function to calculate $J$ between the pairs for hole and electron transport, respectively. Here $J = \langle HOMO^i | \widehat{H} | HOMO^k \rangle$ and $J = \langle LUMO^i | \widehat{H} | LUMO^k \rangle$ for hole and electron transport, respectively. We decompose the reorganization energy $\lambda$ into two parts,

- inner sphere reorganization energy and

- outer sphere reorganization energy.

Inner sphere reorganization energy takes care of the change in nuclear degrees of freedom as charge transfer takes place from dsDNA base $i$ to dsDNA base $j$, which we define as

$$\lambda_{ik}^{int} = U_i^{nC} - U_i^{nN} + U_k^{cN} - U_k^{cC} \qquad (3)$$

where $U_i^{nC}(U_i^{cN})$ is the internal energy of neutral (charged ) base in charged (neutral) state geometry and $U_i^{nN}(U_i^{cC})$ is the internal energy of neutral (charged) base in neutral (charged) state geometry. To calculate the terms appearing in equation 3, we first optimize the geometry of hopping sites for both neutral and charged states using Gaussian09[47]. Then we carry out single point energy calculations with the optimized geometry for these different charged states to obtain various terms involved in the internal reorganization energy appearing in equation 3.

The outer sphere reorganization energy is the part of the reorganization energy that takes into account the reorganization of the environment as the charge transfer takes place. The calculation of external reorganization energy is very involved and intricate[43]. In our calculation, we take the external reorganization as a parameter, rather than calculating it from the QM. The free energy difference $\Delta G_{ik}$ appearing in the rate expression is the contribution from the various sources as described below

$$\Delta G_{ik=} \Delta G_{ik}^{ext} + \Delta G_{ik}^{int} \qquad (4)$$

Here $\Delta G_{ik}^{ext}$ is the contribution of the external electric field, defined as $\Delta G_{ik}^{ext} = F.d_{ik}$, where $F$ is the applied electric field and $d_{ik}$ is the relative position vector between $i^{th}$ and $k^{th}$ bases. In our case this expression simplifies to the following:

$$\Delta G_{ik}^{ext} = \frac{V}{N} \qquad \text{For relative position of } i^{th} \text{ and } k^{th} \text{ bases as shown in fig 2 (a).}$$

$$= 0 \qquad \text{For relative position of } i^{th} \text{ and } k^{th} \text{ bases as shown in fig 2 (b).}$$

$$= -\frac{V}{N} \qquad \text{For relative position of } i^{th} \text{ and } k^{th} \text{ bases as shown in fig 2 (c).}$$

Here V is the applied voltage and N is the number of base-pairs

$\Delta G_{ik}^{int}$ is the contribution in free energy difference due to different internal energies, which can be written as

$$\Delta G_{ik}^{int} = U_i^{cC} - U_i^{nN} + U_k^{cC} - U_k^{nN} \qquad (5)$$

where $U_i^{cC}(U_i^{nN})$ is the internal energy of base $i$ in the charged (neutral) state and geometry. We calculated $\omega_{ik}$ for all hopping pairs and simulate the charge transport dynamics using the kinetic Monte Carlo (MC) method.

We emphasize here that we have only calculated the charge transport rates between the bases, whereas to completely simulate the experimental situation one should also calculate the rates between the base and electrode. We have not explicitly calculated these rates. The explicit calculation scheme for these rates can be found in a article by Rosa Di Felice group[48]. These rates are chosen such that the calculated conductivity becomes independent of these rates (see supplementary section I). Consequently the conductivity that we report here is expected to correspond to the intrinsic conductivity of the DNA.

For kinetic MC , we developed a code[44] with the following algorithm. We assign a unit positive charge (for hole transport) or negative charge (electron transport) to any hopping site $i$. At this point we initialize the time as $t = 0$. If site $i$ has N neighbours, then the waiting time $\tau$ for the charge is calculated according to the relation:

$$\tau = -\omega_i^{-1} \ln(r_1) \qquad [\omega_i = \sum_{k=1}^{N} \omega_{ik}] \qquad (6)$$

and time is updated as $t = t + \tau$ . Here $k$ is the index of the neighbours (the hopping sites to which the charge can hop) coupled to the particular hopping site $i$ and $r_1$ is a uniform random number between 0 to 1. To decide where the charge will hop among N neighbours, we choose the largest $k$ for which $\frac{\sum_k \omega_{ik}}{\omega_i} \leq r_2$. Here $k$ is the index of the neighbours of site $i$ and $r_2$ is another uniform random number between 0 to 1. The above condition will ensure that the side $k$ is selected with probability $\frac{\omega_{ik}}{\omega_i}$ . Next the position of the charge is updated and the above process is repeated. The current through the DNA is calculated as follows:

$$I = e < v > = e < l > / t$$

$< v >$ is the average velocity of the charge and the $< l >$ is the average distance that the charge has moved in time t.

## Results and discussion

All the DFT calculations mentioned in the previous section used the M06-2X/6-311G level of theory as implemented in the Gaussian 09 package. To validate the methodological scheme described in the previous section, we first calculated the resistance of the dsDNA as a function of dsDNA length at an applied voltage of 5 *V*, and the corresponding results are displayed in Figure 3. The Resistance as a function of dsDNA length for 1 V and 6 V applied voltages are shown in Figure S4 of the supplementary information. The resistance increases almost linearly with the length of the dsDNA as has been observed experimentally[28]. This increase of resistance with molecular length is an essential feature of the thermally activated hopping (Marcus-Hush) mechanism. Next we calculate the current at an applied voltage of 5V through all stretched DNA structures starting from the unstretched one for the 5'end1-5'end2 case. The reference voltage of 5V was chosen because the current through the unstretched dsDNA saturates above 5V (see supplementary section II ). We find that the current exhibits a sharp jump (Figure 4(a)) beyond a critical stretching length of 6 Å, as reported by Tao et. al.[28] at which point the conductance drops by 3 orders of magnitude. We note here that to include the effect of solvation of the dsDNA in the conductance calculation, we treated the surrounding water medium using a polarizable continuum model[49] (PCM) with a static dielectric constant 78.3 and dynamic dielectric constant 1.77. To validate the accuracy of this calculation, we also carried out the current calculation using the B3LYP/6-311G level of theory and find similar conductance jump of $4.1 \times 10^{-8}$ A/V rather than $2.8 \times 10^{-8}$ A/V (See Figure S3 in supporting information). It is important to note that without the solvation effect, we do not get the sharp conductance jump shown in figure S3 in the supporting information.

Next we calculate the current as a function of stretching length for the other three pulling protocols using the same level of theory and solvation model as for the 5'end1-5'end2 case presented in the previous section. Here we emphasize the complexity of the current calculation presented in this work. We calculated the current through almost 200 dsDNA structures for the 5'end1-5'end2 case while we calculated the current through 600, 300, 400 dsDNA structures for 3'end1-3'end2, 3'end1-5'end2, 3'end1-5'end1 cases, respectively. The jump in current was seen for all the four cases but at different stretching lengths (Figure 4). Table 1 tabulates the numerical value critical stretching length $l_c$ for the various pulling protocols. A high critical stretching length (32 Å) was found for the 3'end1-3'end2 case, but it was very short (6 Å) for the 5'end1-5'end2 case.

| Pulling Protocol | $l_c$ (Å) |
|---|---|
| 5'end1-5'end2 | 6+/-1 |
| 3'end1-3'end2 | 32+/-5 |
| 3'end1-5'end2 | 23+/-2 |
| 3'end1-5end1 | 11.25+/-2 |

Table 1. Numerical value of the critical stretching length ($l_c$) for various pulling protocols.

To extract a molecular level understanding of the pulling protocol dependent conductance behavior, we now examine the dsDNA structures at various stretching lengths for the various pulling protocols. Since the external environment to the dsDNA does not change during the mechanical pulling, the decay in conductance jump can be ascribed to the structural changes in DNA structures which in turns affect the rate equation. Several instantaneous snapshots of the dsDNA resulting from the pulling simulations are shown in Figure 5. This shows a clear indication of different structures appearing for different pulling cases. As a quantitative measure of the structural changes, we calculate the number of intact hydrogen bonds as a function of the applied force for all four cases (Figure 6 ). In the 5'end1-5'end2 and 3'end1-5'end1 modes, all h-bonds are cleaved within a force ~600 pN, leading to a melted DNA state while 80% h-bond retention is observed for the 3'end1-3'end2 mode, indicating the B-S structural transition of the DNA. The 3'end1-5'end2 pulling case is intermediate of these two extreme situations. To further investigate how the structural change modifies the hopping rate, we first identified the defects (shown in red circle in Figure 7) that appear in the dsDNA during the course of mechanical pulling and plot the transfer integral (Figure 7) between the corresponding bases in the region of defect. We also show the changes in the total number of intact hydrogen bonds (Figure 7) between the base-pairs as the defect in the dsDNA appears for all four pulling protocols. It is clear from Figure 7 that for all the four different pulling protocols, creation of a defect causes the sharp reduction of the transfer integral which in turn affects the charge transfer rate, resulting in the sharp attenuation of the current. The appearance of the defect can be associated with the change in the total number of hydrogen-bonds. The reason behind this dramatic difference in stretching length for the 5'end1-5'end2 pulling and 3'end1-3'end2 pulling is that the structure of the dsDNA stretched from 3'end1-3'end2 ends dramatically differs from the structure stretched form 5'end1-5'end2 ends[33]. This difference in structure arising from 3'end1-3'end2 pulling and 5'end1-5'end2 pulling can also be understood by plotting inclination angle as a function of applied force (see Figure S5 of the supplementary information). Initially the DNA is in B form where the base pairs are tilted with respect to the global axis of the dsDNA[33]. As one stretches the DNA from 5'end1-5'end2 ends, this tilt gets increased which causes the breakage of terminal h-bond resulting the early conductance jump. On the other hand, the 3'end1-3'end2 pulling decreases the base pair tilt. As a result no early breakage of h-bond occurs, rather the DNA undergoes structural transformation from B to S form.

The calculation of the current through the DNA was carried out assuming that the charge transport through the DNA happens via incoherent hopping of charge through the bases. Since the DNA we have studied is short in length (12bp), the transport of charge will have also contribution from tunneling[50, 51]. However the main conclusions (critical stretching length) of the paper remains and do not change due to the inclusion of tunneling (see section X of the supplementary information).

The stretching length vs current calculation described above was at 300 K. To understand the effect of temperature on the dynamics we also calculated the response for 5'end1-5'end2 pulling at 350K and 250K. The h-bond profiles are shown in figure S7(a) in the SI. The dependence of the number of h-bonds as a function of force is similar at all three temperatures but we find faster h-bond decay with force at higher temperature. However, the conductance jump at 6 Å remains largely unaffected by the temperature as shown in figure S7(b) of SI. These results that the decrease in h-bonds with force is faster at high temperatures suggest that there is an activation process involving a barrier of 0.114 eV[52], whereas the temperature independence of the transition at $6\,\mathring{A}$ length, suggests that this may be geometrically related to the stiff bonds of the system.

The high voltage of 5 V was chosen from the experimental V-I characteristics of the DNA, where it is observed that the current saturates at high voltage. At low voltage the current will be low but the behavior of stretching length vs. current plot does not change. To demonstrate this, we report in figure S8 in the SI the current vs stretching length for bias voltage of 1 V and 0.5 V. The critical stretching length does not change at the smaller bias, but of course the current is lower.

Our calculations have assumed that the external reorganization energy is zero . However introducing the external reorganization energy would not affect the behavior of current as a function of end-to-end length of the dsDNA, it only reduces the total magnitude of the current (see supplementary section VI).

All calculations reported in this section are for a specific sequence of the DNA. In order to show that, the structural changes of the dsDNA (which are responsible for the conductance jump) for various pulling protocols do not depend significantly on the specific sequence of the DNA, we carried out studies for two other dramatically different sequences. As shown in Section IX (Fig. S9) of the supplementary Material, the difference sequences lead to the same structural patterns with pulling. Thus we expect the results reported to be independent of the specific DNA sequence.

## SUMMARY AND CONCLUSION

In summary, we have predicted the behaviour of current through dsDNA molecules kept between two electrodes as we stretch the molecule mechanically using four different protocols. Our calculation of current is based on the thermally activated hopping mechanism, which takes into account a very realistic description of DNA structures and of the external environment of the DNA. We find that the response of current through DNA under mechanical pulling depends strongly on the pulling mechanism. We find an abrupt jump in current of almost three order of magnitude within a very short stretching length ($6\,\mathring{A}$) in case of 5'end1-5'end2 pulling while in case of 3'end1-3'end2 pulling it takes almost $32\,\mathring{A}$ of stretching (84%) to observe a similar jump in current . We demonstrate that this change of current is associated with the change of transfer integral $J_{ik}$ which is manifested by the structural changes of DNA.

Thus these calculations provide an atomistic understanding of the behavior of DNA conductance under mechanical tension. These results further explain the piezoresistive behavior[30] of DNA which is an important property for its application to nanodevices. The point at which there is a jump in conductance as a function of stretching will also help in developing a DNA based mechanical switch[53]. We expect these calculations should help further development of DNA nanotechnology.

## Acknowledgements

We thank DST, India for financial support. We thank Dr. Andres Jaramillo-Botero for helpful discussions.

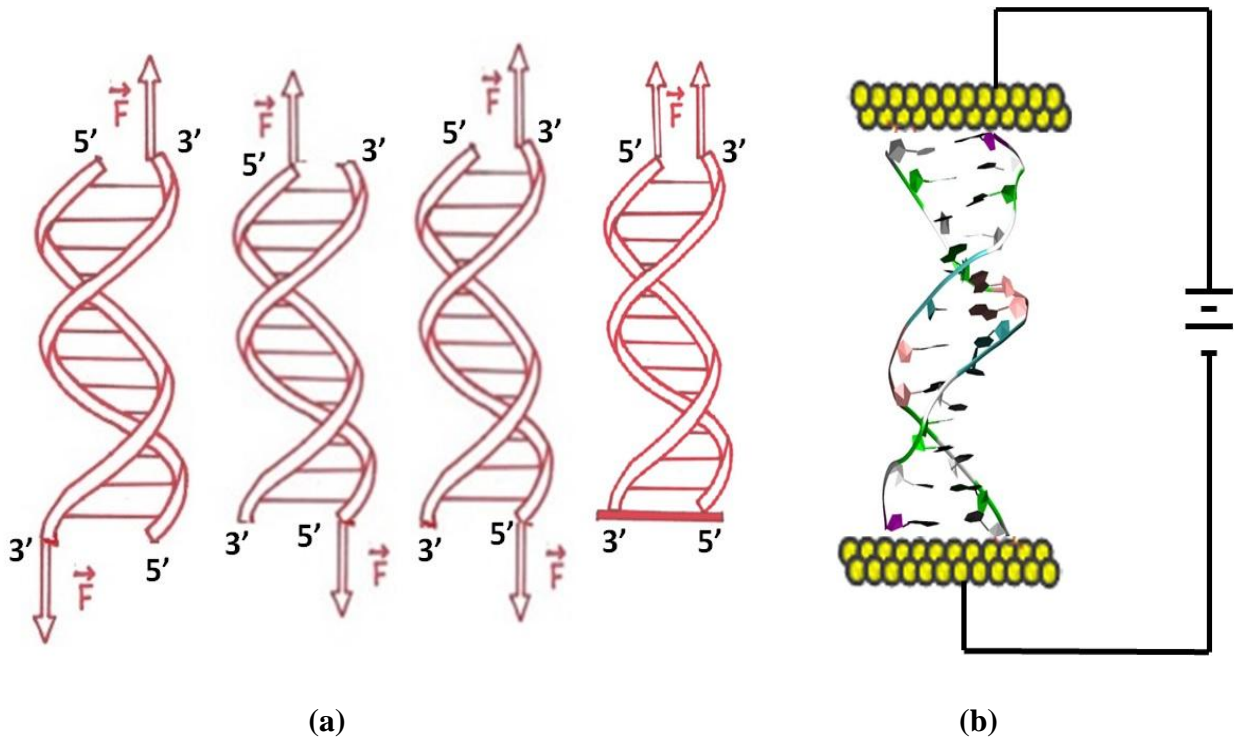

**(a)**                        **(b)**

Figure 1. Schematic diagram of the forcing protocol for applying the force on the DNA. DNA is pulled from 3'end1-3'end2 ends, 5'end1-5'end2 ends, 3'end1-5'end2 ends and 3'end1-5'end1 ends. (b) Schematic diagram of the experimental situation that resembles our simulation.

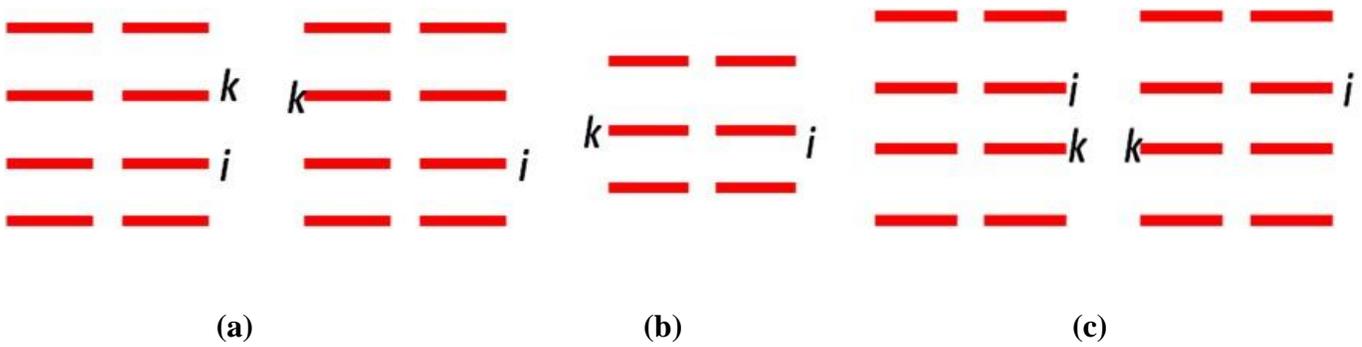

**(a)**  **(b)**  **(c)**

Figure 2. Schematic representation of the relative positions of the charge hopping pairs (DNA bases) in the experimental situation described in Figure 3. If charge is on site , it can hop upwards to site $k$ (a), to the site $k$ in same plane (b) or downwards to the site $k$ (c).

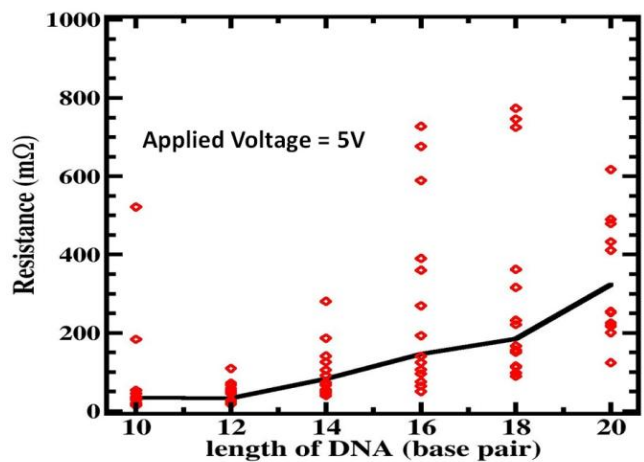

Figure 3. Resistance of the dsDNA with increasing length. The red points indicate the individual case studied and a mean line is drawn to guide the eyes. The resistance increases almost linearly with the length of the DNA. This Increase of resistance with molecular length, which was also observed in the experiment, is an essential feature of the thermally activated hopping (Marcus) mechanism.

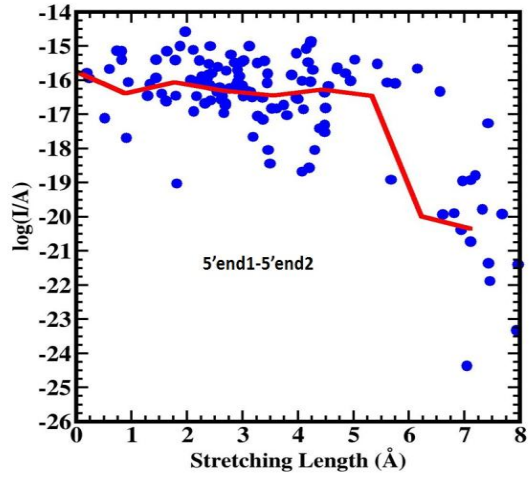

**(a)**

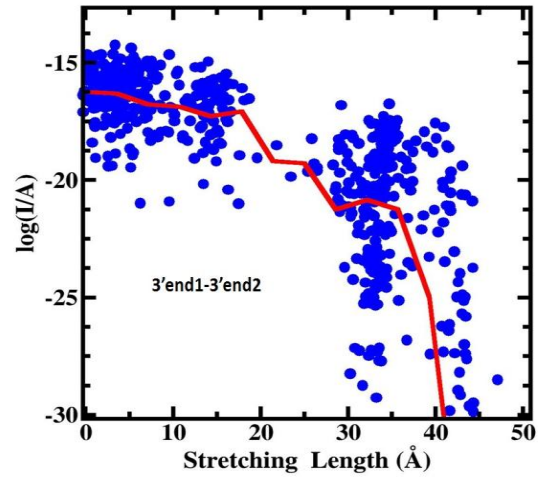

**(b)**

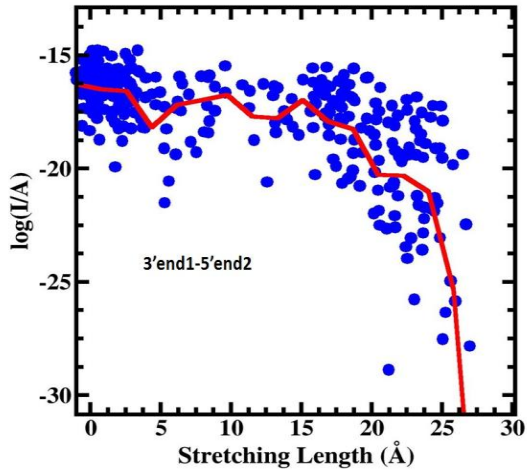

**(c)**

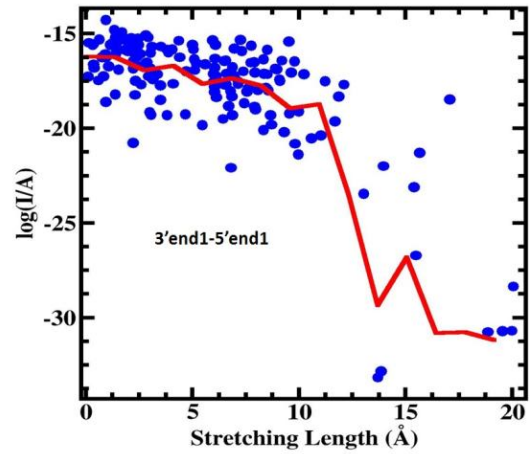

**(d)**

Figure 4. Current through the DNA as a function of the stretching length for pulling from 5'end1-5'end2 (a) ends, 3'end1-3'end2 ends (b), 3'end1-5'end2 (c) and 3'end1-5'end1 (d) . The actual value of current for all individual cases  studied are shown in blue dots. A mean line in red is drawn to guide the eyes.

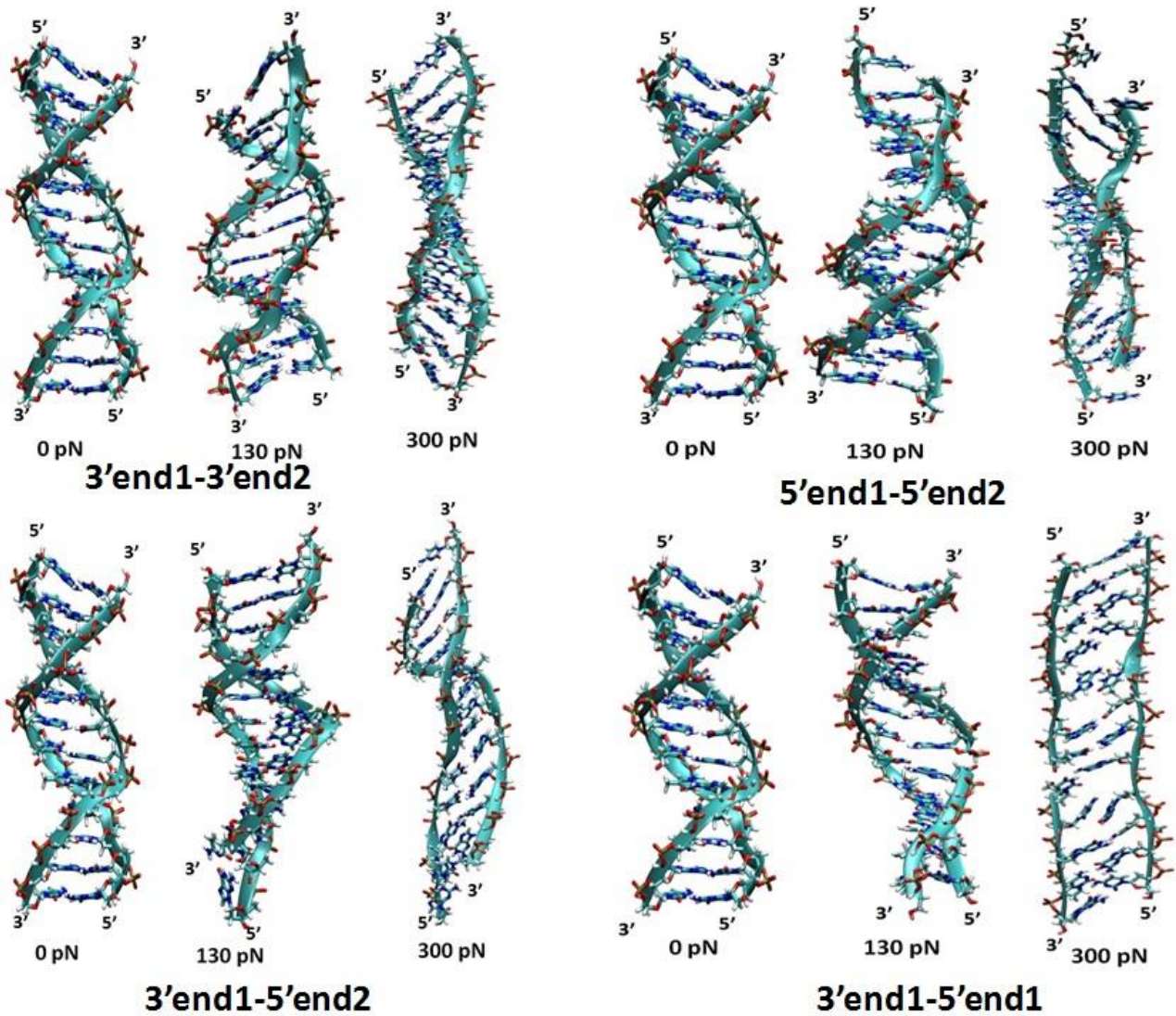

Figure 5. Instantaneous snapshot of the DNA with increasing pulling forces for pulling 3'end1-3'end2 ends, 5'end1-5'end2 ends, 3'end1-5'end2 ends and 3'end1-5'end1 ends respectively.

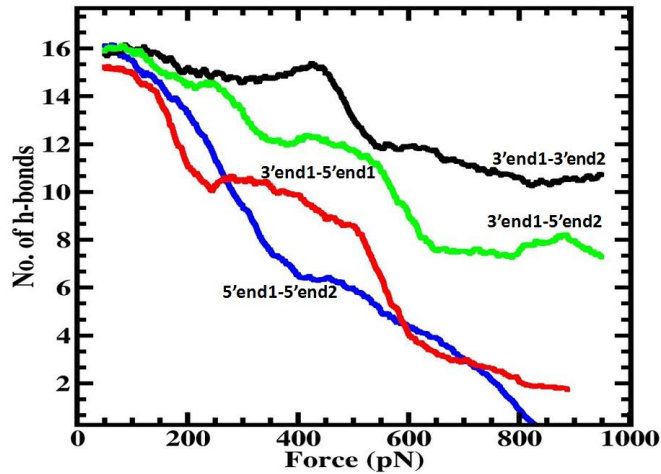

Figure 6: Number of hydrogen bonds as a function of applied force on DNA for 5'end1-5'end2 and 3'end1-3'end2, 3'end1-5'end1 and 3'end1-5'end2 pulling cases. In the case of 5'end1-5'end2 and 3'end1-5'end1 pulling all hydrogen-bonds get cleaved within ~600 pN force indicating a melted DNA state while 80% h-bond remain intact for 3'end1-3'end2 pulling cases which indicates B-S structural transition of the DNA. The 3'end1-5'end2 pulling case is intermediate of this two extreme situation.

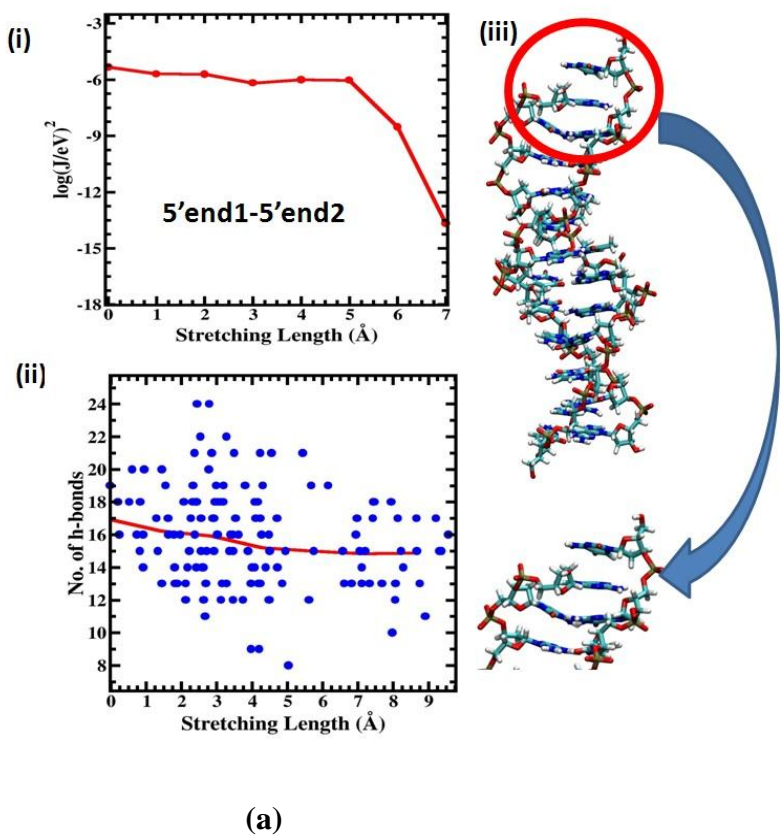

**(a)**

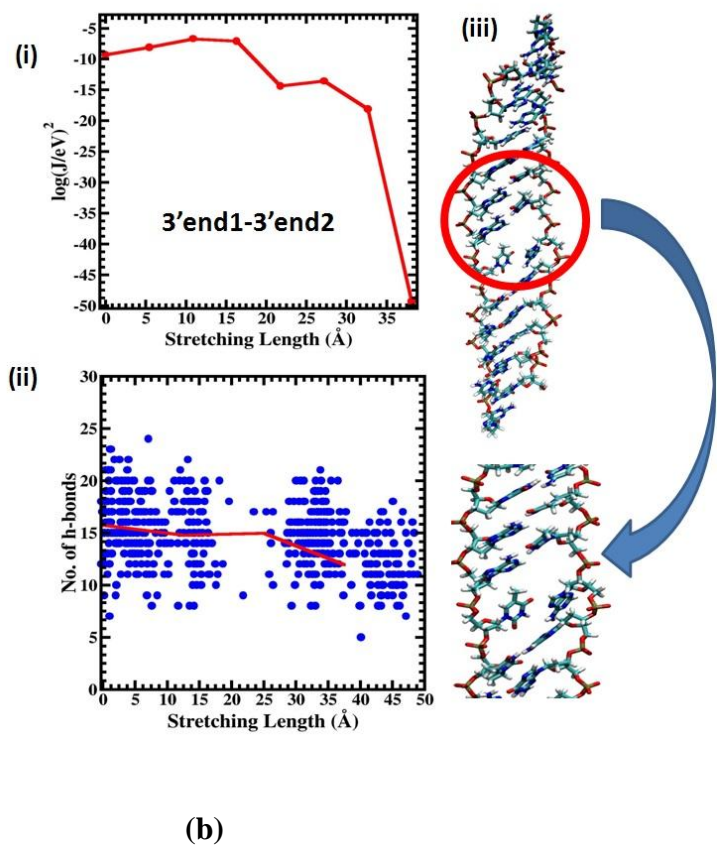

**(b)**

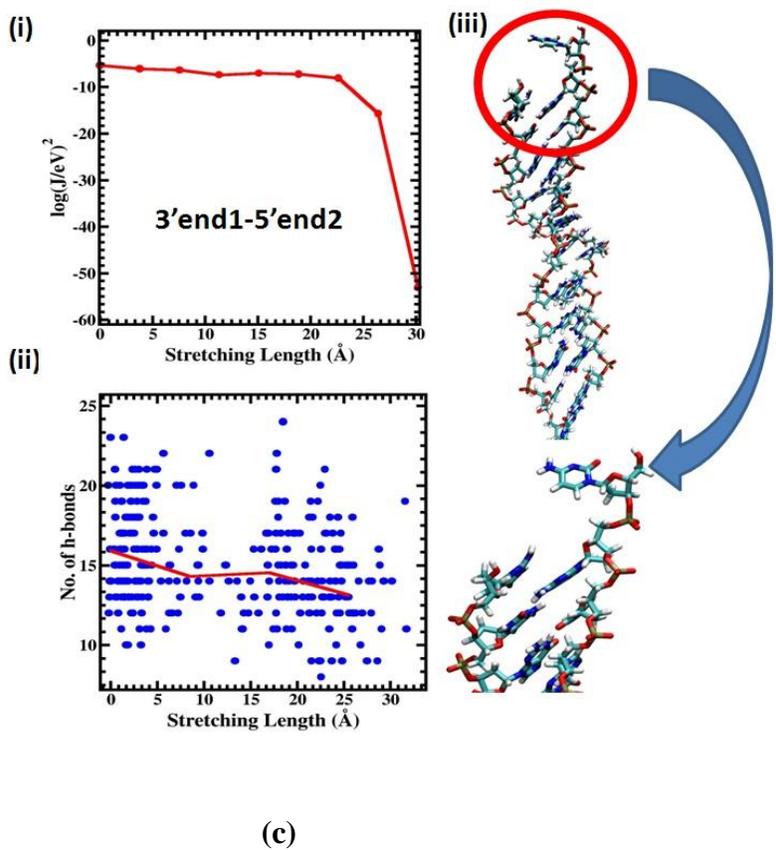

**(c)**

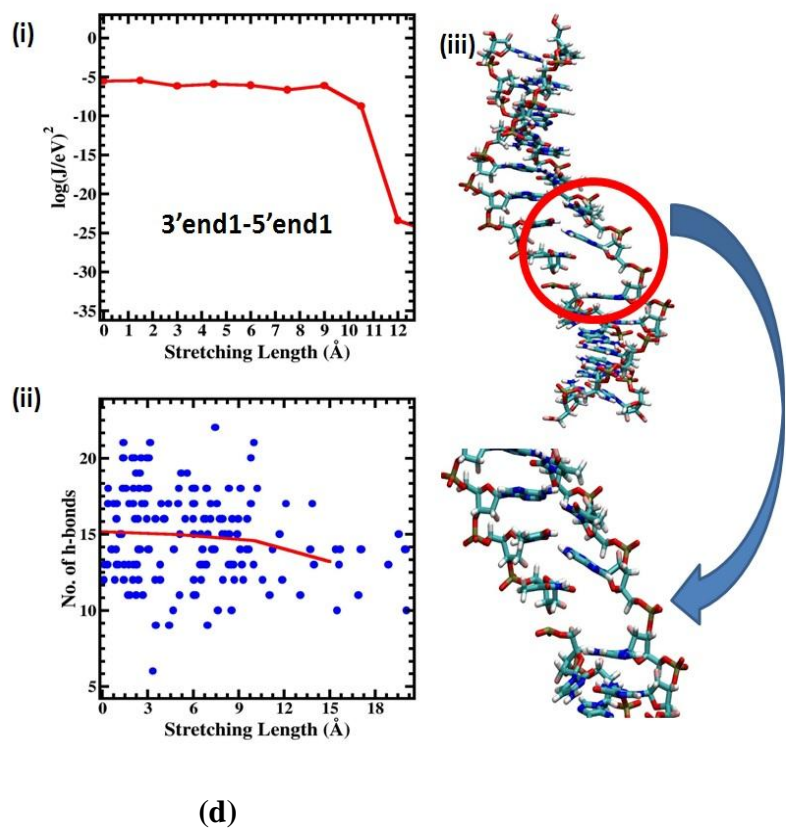

**(d)**

Figure 7.

(i) Transfer integral as a function of stretching length.

(ii) Number of h-bonds as a function of stretching length. Blue dots are the number of individual case and the red line is the average.

 (iii) Appearance of defect in the DNA structure in the course of mechanical pulling

(a) 5'end1-5'end2' case

(b) 3'end1-3'end2 case

(c) 3'end1-5'end2  case

(d) 3'end1-5'end1  case.

Appearance of the defect is associated with the sharp reduction in transfer integral and change in the total no. of h-bonds.